\documentclass[twocolumn,aps, prl, groupedaddress,epsf]{revtex4}

\usepackage{graphicx}

\begin{document}

\title{Electronic transport and quantum Hall effect in bipolar graphene p-n-p junctions}

\author{Barbaros \"{O}zyilmaz$^{1*}$
}
\author{Pablo Jarillo-Herrero$^{1*}$}
\author{Dmitri Efetov$^1$}
\author{Dmitri A. Abanin$^2$}
\author{Leonid S. Levitov$^2$}
\author{Philip Kim$^{1\dag}$}

\affiliation{$^1$Department of Physics, Columbia University, New
York, New York 10027, USA}

\affiliation{$^2$Department of Physics, Center for Materials
Sciences and Engineering, Massachusetts Institute of Technology,
Cambridge, MA 02139}

\begin{abstract}
We have developed a device fabrication process to pattern graphene
into nanostructures of arbitrary shape and control their
electronic properties using local electrostatic gates. Electronic
transport measurements have been used to characterize locally
gated bipolar graphene $p$-$n$-$p$
junctions. We observe a series of fractional
quantum Hall conductance plateaus at high magnetic fields as the
local charge density is varied in the $p$ and $n$ regions. These
fractional plateaus, originating from chiral
edge states equilibration at the $p$-$n$ interfaces,
exhibit sensitivity to inter-edge backscattering
which is found to be strong for some of the plateuas
and much weaker for other plateaus. We use this effect to explore
the role of backscattering and estimate disorder strength
in our graphene devices.
\end{abstract}

\maketitle


Graphene, a recently discovered single sheet of
graphite~\cite{(1)}, stands out as an exceptional material both in
terms of the fundamental physics associated with its unique
"quasi-relativistic" carrier dynamics and potential applications
in electronic devices~\cite{(2)}. Most experiments to date
depended on the presence of a heavily doped Si substrate which
serves as a global back gate, inducing charge density via the
electric field effect. Although such global gate approach yields
interesting transport phenomena~\cite{(1),(3),(4),(5),(6),(7)}, it
represents only the first step towards more complex graphene
devices. 
The use of local gates enables the fabrication of in-plane
graphene heterostructures~\cite{(9-1),(22)}.
A number of applications for local gate devices have been proposed
in order to investigate Klein tunneling~\cite{(10)}, electron
Veselago-lens~\cite{(11)}, quantum point contacts~\cite{(12)} and
quantum dots~\cite{(13),(14)}.

In this letter, we study locally gated graphene devices in the
quantum Hall (QH) regime.
By independently varying voltage on the back gate and
local gate, we can study bipolar QH transport in graphene
$p$-$n$-$p$ heterojunctions in different charge density regimes.
We find a series of fractional QH conductance plateaus as the
local charge density is varied in the $p$ and $n$ regions.
Recently similar QH effect has been reported in a single $p$-$n$
graphene heterojunction\,\cite{(29)}. Our double junction system
allows to study new interesting transport regimes that are absent
in the QH edge transport in a single junction, in particular,
partial equilibration of graphene QH edge states. Conspicuously,
some of our fractionally quantized plateaus are found to be
considerably more fragile with respect to disorder than the other
plateaus. We analyze the distinction in roughness of different
plateaus and show that it points to the importance of inter-edge
backscattering in our narrow graphene samples.

\begin{figure}
\includegraphics[angle=0,width=1.0\linewidth]{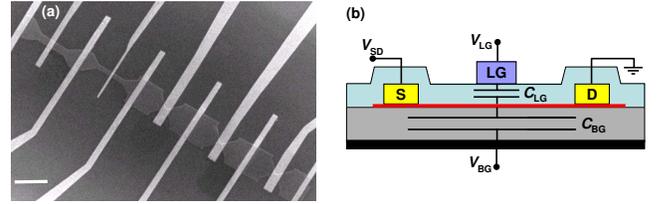}
\caption{(a) Scanning electron microscopy (SEM) picture showing
several complete two-probe devices with local gates. Scale bar
2~$\mu$m. The electrodes contact the widest parts of the
structure. The local gates cover the graphene constrictions and
part of the central graphene bars. (b) Schematic side view of the
devices. Graphene is contacted by source (S) and drain (D)
electrodes, and separated from the back gate plane by 300\,nm
SiO$_2$, and from the local gate (LG) by the top dielectric. The
back gate (with voltage $V_{BG}$) is coupled to the entire
graphene structure via the capacitor $C_{BG}$. The local gate only
couples to part of the structure via $C_{LG}$.}\label{fig1}
\end{figure}

\begin{figure}
\includegraphics[width=1.0\linewidth]{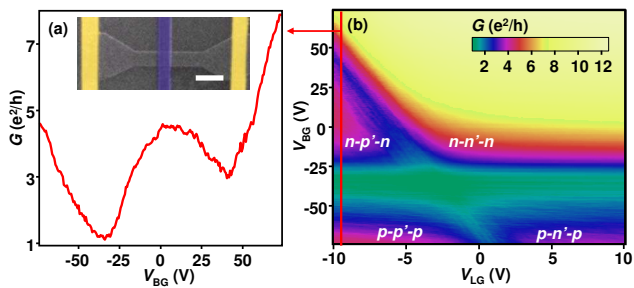}
\caption{(a) $G(V_{BG})$ for a graphene $p$-$n$-$p$ junction,
extracted from (b), showing the two conductance minima associated
with Dirac valleys in the graphene leads and under the local gate.
Inset: false color SEM picture of a patterned graphene bar with
contacts and local gate. Scale bar represents 1~$\mu$m. (b)
Two-dimensional plot of $G(V_{LG}, V_{BG})$ for the device shown
in (a).
}\label{fig2}
\end{figure}

Our devices have been fabricated following a combination of an
etching and dielectric deposition techniques~\cite{(20)}.
Fig.~1(a) shows examples of nanostructured graphene devices after
the complete fabrication process. The devices consist of a
designed graphene nanostructure sandwiched between two
dielectrics, with a global back gate (the highly doped Si
substrate) and partially covered by one or more local gates
(Fig.~1(b)). Such gates allow us to tune the location of the Fermi
energy in graphene globally (via the back gate voltage, $V_{BG}$)
or locally (via the local gate voltage, $V_{LG}$). The
conductance, $G$, of our devices is measured at cryogenic
temperatures (1.5 to 4.2~K), as a function of $V_{BG}$ and
$V_{LG}$, by using a lock-in technique with an ac excitation
voltage of 100~$\mu$V. The single layer character of our devices
is determined by Raman spectroscopy~\cite{(18)} and/or quantum
Hall effect measurements~\cite{(3),(4)}.

Bulk graphene is a zero band gap semiconductor. Therefore, the
Fermi energy in graphene can be continuously varied from valence
to conduction band via the field effect. Incorporating local gates
allows us to induce different charge densities at different sample
regions. Of particular interest is the case when the Fermi energy
in one region is in the valence band ($p$-type) while in the other
region it is in the conduction band ($n$-type).

We have fabricated graphene $p$-$n$-$p$/$n$-$p$-$n$ devices with
different gated channel width and length (see Fig.~1(a)). In total
6 devices in two different single layer graphene pieces were
studied and found to exhibit similar characteristics. Fig.~2(b)
shows $G (V_{LG}, V_{BG})$ at zero magnetic field for a typical
device~\cite{(22)}. The most prominent feature is the presence of
two conductance minima valleys: one approximately horizontal,
independent of the voltage $V_{LG}$, and the other diagonal. The
first valley tracks the charge neutrality point (or, Dirac point)
in the regions outside the local gate, further denoted as graphene
leads (GLs).The second valley, whose position is controlled both
by the local and the back gate voltage, tracks the neutrality
point in the region under the local gate (LGR). The slope of this
valley is equal to the ratio of the local capacitance to the back
gate capacitance. For the device shown in Fig.~2(b) this ratio is
10.5.
Note that the typical local gate breakdown voltage for our
dielectric heterostructure is larger than $12\,{\rm V}$, which,
together with a ten-fold enhancement of the gate coupling factor,
means that we can induce local charge densities at least
comparable, and often larger, than with the back gate (where
leakage typically starts to occur at $V_{BG}\sim 100\,{\rm V}$).

Away from the Dirac valleys, the conductance increases with
increasing charge density (see Fig.\,2). The two valleys separate
four regions in the plot: $p$-$p'$-$p$, $p$-$n'$-$p$, $n$-$n'$-$n$
and $n$-$p'$-$n$, where $n$/$p$ refers to negative/positive charge
density and the prime indicates density in the LGR. The
conductance is not symmetric across the valleys, because for
opposite polarities there is an extra contribution due to the
resistance of the two $p$-$n$ interfaces~\cite{(22)}. We also note
here that even in the $p$-$n$-$p$ and $n$-$p$-$n$ regions, the
device shows considerable conductance ($G>e^2/h$) without any
signature of rectifying behavior, as expected for transport in a
zero-gap heterojunction. In fact, graphene is the only
two-dimensional electron gas (2DEG) where in-plane bipolar
heterostructures $p$-$n'$-$p$ and $n$-$p'$-$n$ can be studied in
the linear response regime.

The lateral graphene heterojunctions exhibit interesting phenomena
at high magnetic field. One of the hallmarks of graphene is the
relativistic integer Quantum Hall (QH) effect, manifested in a
series of conductance plateaus at half-integer multiples of
$4e^2/h$~\cite{(3),(4),(23)}. Such unique QH plateau structure can
be attributed to an odd number of QH edge states that carry
current with conductance $2e^2/h$~\cite{(24),(25)}. The capability
of placing local tunable electrostatic barriers/wells along the
current pathway allows us to use QH mode propagation to explore
intrinsic transport characteristics of graphene heterojunction
structures.

\begin{figure}
\includegraphics[width=1.0\linewidth]{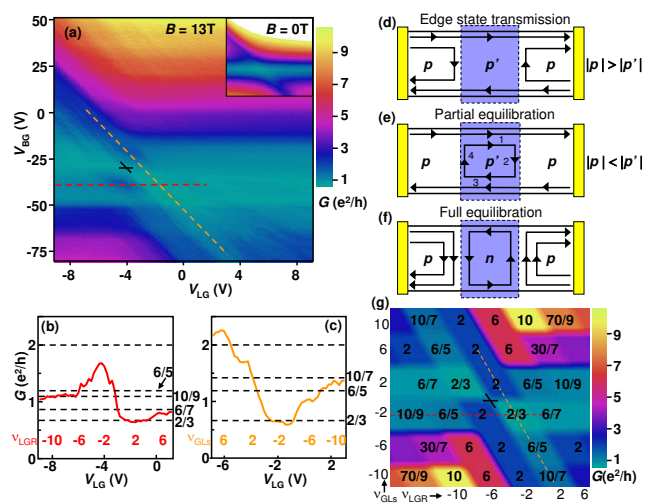}
\caption{(a) Color map of conductance $G(V_{LG},V_{BG})$ at
magnetic field $B = 13$~T, and $T = 4.2$~K. The black cross
indicates the location of filling factor zero in LGR and GLs.
Inset: Conductance at zero $B$ in the same $(V_{LG},V_{BG})$ range
and the same color scale as main figure (white denotes $G>10.5
e^2/h$). (b) $G(V_{LG})$ extracted from (a), red trace, showing
fractional values of the conductance. Numbers on the right
indicate expected fractions for the various filling factors (red
numbers indicate the filling factor, $\nu'$, in LGR), see also
(g). (c) $G(V_{LG})$ (projection of orange trace from (a) onto
$V_{LG}$-axis). Orange numbers indicate filling factor, $\nu $, in
the GLs. (d) to (f): different edge state diagrams representing
possible equilibration processes taking place at different charge
densities in the GLs and LGR. The purple region indicates the LGR.
Yellow boxes indicate contact electrodes. (g) Simulated color map
of the theoretical conductance plateaus expected from the
mechanisms shown in (d-f) for different filling factors in the GLs
and LGR. The numbers in the rhombi indicate the conductance at
that plateau.
The color scale is identical to that of (a).} \label{fig3}
\end{figure}

Fig.\ref{fig3}a shows $G(V_{LG}, V_{BG})$ at high magnetic field,
$B$, for a typical device. In addition to the four regions seen at
zero field (Fig.\ref{fig3}a, inset), the conductance map reveals a
rich pattern consisting of rhombi and bands where $G$ exhibits
plateaus. Overall this pattern is symmetric with respect to the
neutrality point (marked with a black cross) which corresponds to
$\nu =\nu'=0$. (Here $\nu $ and $\nu'$ are the Landau level
filling factors, equal to $n_ce/hB$ with $n_c$ the carrier density
in GLs and LGR, respectively.) While the plateaus in conductance
at high densities $|\nu |,\,|\nu'|\geq 6$, are well accounted for
by a two resistors-in-series model, with each resistor
corresponding to the QH conductance of GLs, a more complex and
interesting behavior is observed at lower densities, where
resistors cease to add up in a classical fashion.

The nonclassical behavior is found in particular at low filling
factors, especially when either $\nu'$ or $\nu $ equals $+2$ or
$-2$ (red and orange traces in Fig.\ref{fig3}(a)). Notably, we
observe conductance plateaus at values close to fractional values
of the conductance quantum, $e^2/h$. Such simple fractions
include,  for example, $(2/3)e^2/h$, $(6/7)e^2/h$ and
$(10/9)e^2/h$ (Fig.\ref{fig3}(b)). These values are in sharp
contrast to the conductance plateaus at $(2,6,10...)e^2/h$,
observed in homogenous two-terminal devices~\cite{(7)}.

The unusual fractional conductance plateau patterns can be
analyzed by using models developed for QHE mode propagation in
2DEGs with density gradients~\cite{(26),(27)}. Our graphene
system, however, represents a distinct advantage owing to the
possibility of creating opposite polarities of charge carriers in
adjacent regions.

The simplest case to consider is when the polarity of GLs and LGR
is the same (either $n$-$n'$-$n$ or $p$-$p'$-$p$), and the LGR
density is lower than the GLs density: $|\nu'|\le |\nu|$. In this
case, as shown in Fig.\ref{fig3}(d), the number of QH edge modes
is larger in the GLs than in the LGR. The modes existing only in
the GLs are fully reflected at the  LGR-GLs
interfaces~\cite{(26),(27)}, while those present in both regions
exhibit full transmission, giving rise to the net conductance
$G=(e^2/h)|\nu'|$.

A more interesting situation occurs when the LGR density is higher
than the GLs density, with the LGR and GLs polarities still the
same. In this case the number of edge states is smaller in the GLs
than in the LGR (see Fig.\ref{fig3}(e)). Crucially, the states
circulating in LGR can produce partial equilibration among the
different channels, because they couple modes with different
electrochemical potentials. To analyze this regime, we suppose
that current $I$ is injected from the left lead, while no current
is injected from the right lead. Then the conservation of current
yields $I+I_4=I_1$, $I_2=I_3$ (the LGR edges are labeled by 1, 2,
3, 4 as shown on Fig.\ref{fig3}(e)). Assuming that the current at
the upper and lower LGR edges is partitioned equally among
available edge modes, we obtain the relations for the current
flowing out of these edges: $I_2=rI_1$, $I_4=rI_3$,
($r=1-\nu/\nu'$). Solving these equations for $I_{1...4}$, we
determine the current flowing in the drain lead as
$I_{out}=I_1-I_2$ and find the net conductance
\begin{equation}
\label{eq:partial_mixing}
G=\frac{e^2}{h}\frac{|\nu'| |\nu |}{2|\nu'|-|\nu |}
=\frac65,\,\frac{10}9,\,\frac{30}7,...
\quad (|\nu'|\ge|\nu|),
\end{equation}
where $\nu,\nu'=\pm2,\pm6...$.
We emphasize that
this {\it partial equilibration} regime can only occur in the
presence of two $n$-$n'$ or $p$-$p'$ interfaces, and would not
occur in a single $n$-$n'$ or $p$-$p'$ junction~\cite{(28),(29)}.

The last, but most unique case is when the GLs and LGR have
opposite carrier polarity. In this case, the edge states
counter-circulate in the $p$ and $n$ areas, running parallel to
each other along the $p$-$n$ interface (see Fig.\ref{fig3}(f)).
Such propagation, leading to mixing among edge states, results in
full equilibration at the $p$-$n$ interfaces:$I_1=rI_2$,
$I_3=rI_4$, ($r=|\nu'|/(|\nu|+|\nu'|)$). Combining this with
current conservation, in this case written as $I+I_1=I_4$,
$I_2=I_3$, we find the currents and obtain the conductance
\begin{equation}
\label{eq:full_mixing}
G=\frac{e^2}{h}\frac{|\nu'| |\nu |}{2|\nu'|+|\nu |}
=\frac23,\,\frac65,\,\frac67,...
\quad (\nu \nu'<0),
\end{equation}
where $\nu,\nu'=\pm2,\pm6...$. The net conductance in this case is
described by three quantum resistors in series.

The summary of all possible conductance values for these three
regimes is shown as a color map in Fig.\ref{fig3}g. Our first
observation is that the structure of the experimental pattern
resembles qualitatively the theoretical one when the filling
factor equals $\pm 2$ either in the GLs or in the LGR. For a
quantitative analysis, we choose two cuts extracted from
Fig.\ref{fig3}(a), showing conductance for fixed $\nu =-2$
(Fig.\ref{fig3}b) and $\nu'=2$ (Fig.\ref{fig3}c). We record
reasonably good plateaus at $G=(2/3)e^2/h$, $G=(10/9)e^2/h$ as
well as other fractions discussed above, with the only exception
of a considerably more poor plateau with $G=2e^2/h$ (see below).
Of particular interest is the non-monotonic conductance behavior
in Fig.\ref{fig3}(b) for $\nu'=2,-2,-6,-10$ (with $\nu =-2$),
which reflects the full equilibration $\rightarrow$ edge state
transmission $\rightarrow$ partial equilibration sequence. This is
in contrast with the monotonic behavior of $G$ in
Fig.\ref{fig3}(c) for $\nu =-2\to 2\to 6$ (with $\nu'=2$), where
only the full equilibration and full transmission regimes are
expected.

\begin{figure}
\includegraphics[angle=0,width=1.0\linewidth]{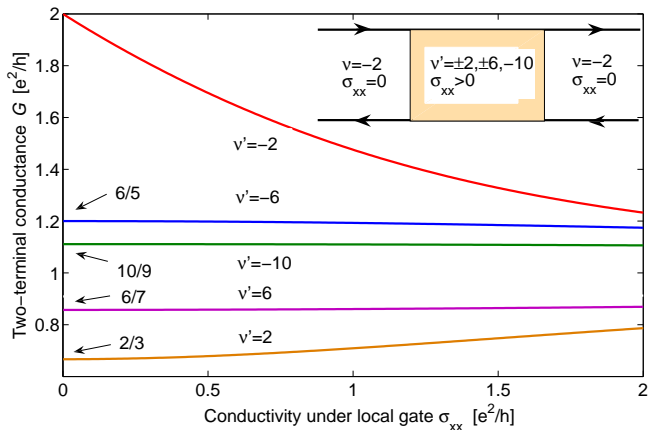}
\caption{Two-terminal conductance of a lateral heterojunction in a
QH state (see inset) as a function of the longitudinal
conductivity under local gate (LGR) for the states with $\nu=-2$
and $\nu'=\pm 2,\pm 6,-10$ (red trace in Fig.\ref{fig3}(a,b)).
Results obtained from the 2D transport
model\,\cite{Girvin81,Ilan2006,(27-1)} for LGR of size $500\,{\rm
nm}\times 700\,{\rm nm}$ are shown. Finite $\sigma_{xx}$ has
considerable effect on the conductance of the state with
$\nu',\nu=-2$, but very little effect on other states. This
explains the difference in roughness of the observed plateaus for
$\nu'\ne-2$ and $\nu'=-2$. } \label{fig4}
\end{figure}

We have measured in total four devices which all exhibit similar
conductance patterns. Notably, in none of these devices $G$
reached the full $2e^2/h$ value at $\nu',\nu =\pm2$, whereas other
conductance plateaus were well developed. The lack of quantization
points to the presence of backscattering between opposite edges of
our sample, which may occur in LGR bulk or in the transitional
regions at the LGR-GLs junctions.

Why are the $\nu',\nu =\pm2$ plateaus so sensitive to
backscattering, while other plateaus are not? To gain insight into
this question, we now investigate how robust are the results
(\ref{eq:partial_mixing}),(\ref{eq:full_mixing}) with respect to
bulk conduction in our QH system. To that end, we consider a 2D
transport model describing the system by local conductivity. Here
we focus on the simplest situation, taking the longitudinal
conductivity $\sigma_{xx}$ nonzero in the gated region (LGR) and
zero outside (GLs), and the Hall conductivity $\sigma_{xy}$ equal
to $\nu'e^2/h$ ($\nu e^2/h$) for LGR (GLs).

An exact solution for 2D current and potential distribution for
this problem was obtained\,\cite{(27-1)} by adapting the conformal
mapping technique developed in Refs.\cite{Girvin81,Ilan2006}. The
resulting two-terminal conductance $G$ of the fractionally
quantized states from the $\nu=-2$ trace (Fig.\ref{fig3}b) are
displayed in Fig.\ref{fig4}. First, we note that the limiting
values of $G$ at $\sigma_{xx}\to 0$ agree with the simple
fractions (\ref{eq:partial_mixing}),(\ref{eq:full_mixing}) derived
above. Furthermore, the effect of finite $\sigma_{xx}$ is
considerably stronger for the $\nu=\nu'=-2$ state than for all
other states --- it is linear rather than quadratic at small
$\sigma_{xx}$. Comparing to the deviation from the quantized value
in Fig.\ref{fig3}, we estimate $\sigma_{xx}\lesssim 0.5 e^2/h$.

We thus infer that weak backscattering is non-detrimental for all
the states except $\nu=\nu'$, which is in agreement with the
observed stability of fractional plateaus. This conclusion also
agrees with the general intuition that current paths in a QH
system are constrained stronger when density is varying in space
than when it is constant\,\cite{Pan95}. We therefore believe that
our understanding of stability of the observed fractional plateaus
is quite generic and insensitive to whether the backscattering in
our graphene devices occurs mainly in LGR bulk or at the LGR-GLs
interfaces.

We thank A. Yacoby, C. Marcus, E. Poortere and K. Bolotin for
stimulating discussions, Inanc Meric for help with the ALD system,
and Yang Wu for the Raman characterization. This work is supported
by the ONR (N000150610138), FENA, NSF CAREER (DMR-0349232) and
NSEC (CHE-0117752), and the New York State Office of Science,
Technology, and Academic Research (NYSTAR). DA and LL
acknowledge support by NSF MRSEC (DMR 02132802) and NSF-NIRT
DMR-0304019.

$^*$These authors contributed equally to this work

$^\dag$E-mail:pk2015@columbia.edu


\begin{references}

\bibitem{(1)} K. S. Novoselov {\it et. al.}, Science {\bf 306}, 666 (2004).

\bibitem{(2)} A. K. Geim, K. S. Novoselov, Nat. Mat. {\bf 6}, 183 (2007).

\bibitem{(3)} K. S. Novoselov {\it et. al.}, Nature {\bf 438}, 197 (2005).

\bibitem{(4)} Y. Zhang, Y. W. Tan, H. L. Stormer, P. Kim, Nature {\bf 438}, 201 (2005).

\bibitem{(5)} K. S. Novoselov {\it et. al.}, Nat. Phys. {\bf 2}, 177 (2006).

\bibitem{(6)} Y. Zhang {\it et. al.}, Phys. Rev. Lett. {\bf 96}, 136806 (2006).

\bibitem{(7)} H. B. Heersche, P. Jarillo-Herrero, J. B. Oostinga, L. M. K. Vandersypen, A. F. Morpurgo, Nature {\bf 446}, 56 (2007).



\bibitem{(9-1)} M. C. Lemme, T. J. Echtermeyer, M. Baus, H. Kurz. IEEE Elec. Dev. Lett. {\bf 28}, 282 (2007).

\bibitem{(10)} V. V. Cheianov, V. I. Fal'ko, Phys. Rev. B {\bf 74}, 041403 (2006).

\bibitem{(11)} V. V. Cheianov, V. Fal'ko, B. L. Altshuler, Science {\bf 315}, 1252
(2007).

\bibitem{(12)} A. Rycerz, J. Tworzydlo, C. W. J. Beenakker, Nat. Phys. {\bf 3},
172 (2007).

\bibitem{(13)} P. G. Silvestrov, K. B. Efetov, Phys. Rev. Lett. {\bf 98}, 016802
(2007).

\bibitem{(14)}B. Trauzettel, D. V. Bulaev, D. Loss, G. Burkard, Nat. Phys.
{\bf 3}, 192 (2007).

\bibitem{(22)} B. Huard,
J. A. Sulpizio, N. Stander, K. Todd, B. Yang, and D. Goldhaber-Gordon,
Phys. Rev. Lett. {\bf 98}, 236803 (2007).

\bibitem{(29)} J. R. Williams, L. C. DiCarlo, C. M. Marcus, cond-mat/0704.3487.

\bibitem{(20)}B. \"{O}zyilmaz {\it et. al.}, cond-mat/0705.3044. During the preparation of this manuscript, we became aware of
related work on graphene $p$-$n$ junctions \cite{(22),(29),(28)}.

\bibitem{(18)}A. C. Ferrari {\it et al.}, Phys. Rev. Lett. {\bf 97}, 187401 (2006).


\bibitem{(28)} D. A. Abanin, L. S. Levitov, cond-mat/0704.3608.

\bibitem{(23)} V. P. Gusynin, S. G. Sharapov, Phys. Rev. Lett. {\bf 95}, 146801
(2005).

\bibitem{(24)} N. M. R. Peres, F. Guinea, A. H. C. Neto, Phys. Rev. B {\bf 73},
125411 (2006).

\bibitem{(25)} D. A. Abanin, P. A. Lee, L. S. Levitov, Phys. Rev. Lett. {\bf 96},
176803 (2006).

\bibitem{(26)} R. J. Haug, A. H. MacDonald, P. Streda, K. von Klitzing,
Phys. Rev. Lett. {\bf 61}, 2797 (1988).

\bibitem{(27)} R. J. Haug, Sem. Sci. Tech. {\bf 8}, 131 (1993).

\bibitem{(27-1)}D. A. Abanin, L. S. Levitov, unpublished.

\bibitem{Girvin81}
R. W. Rendell and S. M. Girvin,
Phys. Rev. B {\bf 23}, 6610 (1981).

\bibitem{Ilan2006}
R. Ilan, N. R. Cooper, and A. Stern,
Phys. Rev. B {\bf 73}, 235333 (2006).

\bibitem{Pan95}
W. Pan {\it et al.},
Phys. Rev. Lett. {\bf 95}, 066808 (2005).

\end{references}
\end{document}